\documentclass[acmtog,nonacm,screen]{acmart}

  \usepackage{booktabs}
  \usepackage{xspace}

  \usepackage[ruled]{algorithm2e}
  
  \SetAlFnt{\small}
  \SetAlCapFnt{\small}
  \SetAlCapNameFnt{\small}
  \SetAlCapHSkip{0pt}

  \newcommand{\sysname}{\textit{MVWeaver}\xspace}

\begin{document}

  \title{MVWeaver: A Hierarchical Music Video Generation Agent
  with a Learned Song-to-Visual Bridge}


\author{Sifei Li}
\authornote{Part of this work was done during an internship at KlingAI.}
  \affiliation{%
    \institution{Institute of Automation, Chinese Academy of Sciences}
    \city{Beijing}
    \country{China}
  }
  \affiliation{%
    \institution{School of Artificial Intelligence,
      University of Chinese Academy of Sciences}
    \city{Beijing}
    \country{China}
  }
  \affiliation{%
    \institution{KlingAI Research}
    \city{Beijing}
    \country{China}
  }
  \email{lisifei2022@ia.ac.cn}

  \author{Minyan Luo}
  \affiliation{%
    \institution{Institute of Automation, Chinese Academy of Sciences}
    \city{Beijing}
    \country{China}
  }
  \affiliation{%
    \institution{School of Artificial Intelligence,
      University of Chinese Academy of Sciences}
    \city{Beijing}
    \country{China}
  }
  \email{luominyan2025@ia.ac.cn}

  \author{Xu Li}
  \authornote{Corresponding authors.}
  \affiliation{%
    \institution{KlingAI Research}
    \city{Beijing}
    \country{China}
  }
  \email{xuliustc1306@gmail.com}

  \author{Guodong Qi}
  \affiliation{%
    \institution{KlingAI Research}
    \city{Beijing}
    \country{China}
  }
  \email{qiguodong@kuaishou.com}

  \author{Xincan Wang}
  \affiliation{%
    \institution{Shanghai Theatre Academy}
    \city{Shanghai}
    \country{China}
  }
  \email{wangxincan@sta.edu.cn}

  \author{Hanwen Wang}
  \affiliation{%
    \department{Department of Directing}
    \institution{Beijing Film Academy}
    \city{Beijing}
    \country{China}
  }
  \email{Wang_Hanwen1993@sina.com}

  \author{Chen Zhang}
  \affiliation{%
    \institution{KlingAI Research}
    \city{Beijing}
    \country{China}
  }
  \email{zhangchen03@kuaishou.com}

  \author{Pengfei Wan}
  \affiliation{%
    \institution{KlingAI Research}
    \city{Beijing}
    \country{China}
  }
  \email{wanpengfei@kuaishou.com}

  \author{Oliver Deussen}
  \affiliation{%
    \institution{University of Konstanz}
    \city{Konstanz}
    \country{Germany}
  }
  \email{oliver.deussen@uni-konstanz.de}

  \author{Weiming Dong}
  \authornotemark[2]
  \affiliation{%
    \institution{Institute of Automation, Chinese Academy of Sciences}
    \city{Beijing}
    \country{China}
  }
  \affiliation{%
    \institution{School of Artificial Intelligence,
      University of Chinese Academy of Sciences}
    \city{Beijing}
    \country{China}
  }
  \email{weiming.dong@ia.ac.cn}

  \renewcommand{\shortauthors}{Sifei Li et al.}

  \begin{abstract}
Music videos are an important form of audiovisual expression in contemporary culture. They translate and extend the expressive content of songs through deliberate visual design. Existing automatic music video (MV) generation systems can generate visually plausible shots, yet often struggle with long-form coherence and song-grounded visual development. We present MVWeaver, a music video generation agent that integrates hierarchical planning with a learned song-to-visual bridge that translates song understanding into executable shot plans. The MVWeaver architecture comprises a comprehensive song analysis module, a visual planner that constructs hierarchical plans, and downstream image and video generation models that render the planned content. To equip a general-purpose LLM with MV-specific song-to-visual knowledge, we learn a bridge between song analysis and visual planning from real-MV-derived supervision and curate 1,861 real-world song--MV pairs with structured song-side, MV-side, and teacher-inferred song-to-visual rationale annotations. Using these annotations, we perform LoRA-based supervised fine-tuning (SFT) of a large language model to predict song-to-visual bridges that guide hierarchical visual planning. Our experiments demonstrate stronger song-grounded visual translation, richer visual development, and greater conceptual and shot-to-shot coherence, while ablations support the benefits of learned bridge conditioning.
\end{abstract}


  \maketitle

  \section{Introduction}
\label{sec:introduction}

If a song conjures a world through sound, then a music video brings that world into view. Progress in controllable song generation~\cite{wang2026segtune}
  further motivates tools for automatically creating accompanying
  music videos. Professional MVs do more than literally illustrate a song, enriching and extending its expression through associated situations, motifs, and visual metaphors~\cite{gow1994mood}. 
Motivated by conceptual blending~\cite{fauconnier1998conceptual}, we view MV planning as deriving plausible visual associations from song evidence and developing them into a coherent MV plan.
Recent advances in image and video generation make automatic full-song MV creation increasingly feasible.
Given a song, the task is to analyze it, develop a coherent visual
  concept, and generate a full-length music video that is musically
  aligned and visually consistent.

Research relevant to automatic MV creation has expanded beyond clip-level synthesis to encompass agentic multi-shot production, music-aware editing, and end-to-end MV generation. (1) Agentic video systems, including AniME~\cite{zhang2025anime}, AniMaker~\cite{shi2025animaker}, VISTA~\cite{long2026vista}, and STAGE~\cite{zhang2026stage}, automate long or multi-shot workflows and improve continuity, but are driven primarily by stories or text prompts rather than creating a song-to-visual translation. (2) Music-guided editing methods such as GLANCE~\cite{lin2026glance} and BEAT~\cite{wang2026beat} align and arrange existing footage according to musical structure, without constructing a new, original visual world from the song. (3) Beyond video editing, recent work explores generating MVs directly from songs. YingVideo-MV~\cite{chen2025yingvideo} combines music-aware shot planning with a dedicated performance-video generator, emphasizing controllable camera motion and audio-driven lip synchronization. AutoMV~\cite{tang2025automv} is a training-free multi-agent pipeline that coordinates music preprocessing, screenwriting, directing, a character bank, heterogeneous generators, and verification for full-song production. AllocMV~\cite{wang2026allocmv} targets cost-aware long-horizon synthesis through persistent character and scene states, saliency-based resource allocation, and visual-prefix reuse. However, their visual planning relies largely on general-purpose LLMs rather than domain knowledge distilled from the song-to-visual associations embodied in real MVs. Creator-facing MV platforms such as VidMuse\footnote{\url{https://vidmuse.ai/zh-CN}} and Seko\footnote{\url{https://seko.sensetime.com/explore}} rely heavily on user-provided materials and creative guidance to achieve high-quality outputs.

In this paper, we propose \sysname, which couples a hierarchical production agent with a learned song-to-visual bridge. The agent integrates comprehensive song analysis, multi-level visual planning, and downstream generation. Its planner separates conceptual development from executable realization by progressively expanding an MV brief into ordered visual developments, scenes, and shots, while propagating reusable assets across levels and aligning scene and shot boundaries to candidate rhythmic points. Global and local bridge decisions condition the MV-brief and visual-development stages, respectively, shaping the overall visual premise and its subsequent development before clip generation. We trained our bridge on 1,841 song-MV pairs using teacher-inferred rationales that connect song evidence to visual choices. Rather than supervising a direct jump to MV choices, these rationales expose the intermediate associations underlying visual decisions and provide song-to-visual knowledge distilled from real MVs.

\begin{figure*}[t]
    \centering
    \includegraphics[width=\textwidth]{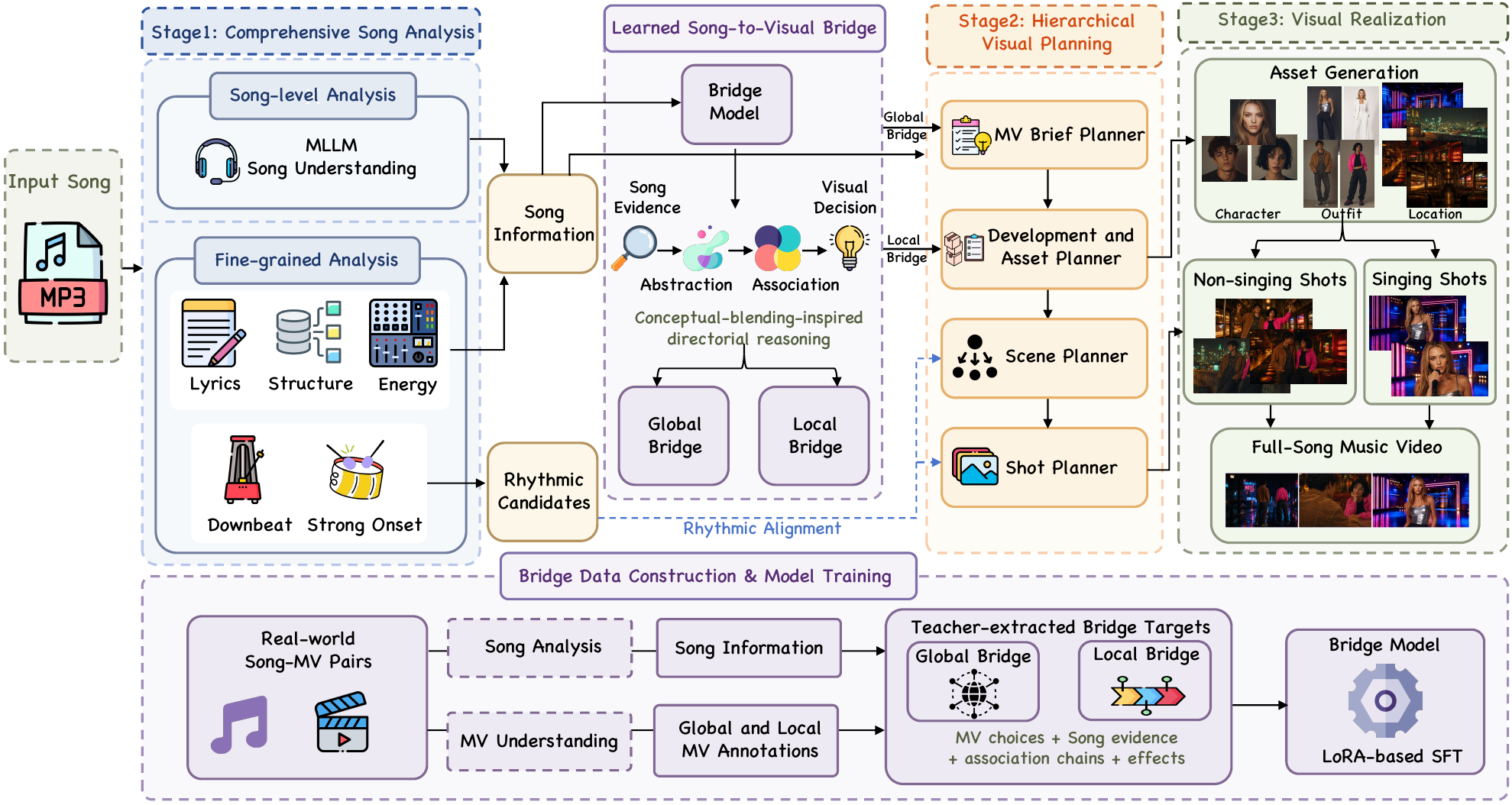}
    \caption{
    Overview of \sysname. The pipeline is organized into a bridge-learning branch and a bridge-guided MV generation branch. The former derives supervision from real-world song--MV pairs for LoRA-based SFT. In the latter, comprehensive song analysis produces global and local bridge decisions, which are progressively expanded through hierarchical visual planning and rendered by reference-conditioned image and video generators.
    }
    \label{fig:mvweaver_pipeline}
\end{figure*}

Our contributions can be summarized as follows:
\begin{itemize}
    \item \sysname, an end-to-end hierarchical agent that connects comprehensive song analysis, multi-level visual planning, and downstream generation for full-song MVs.
    \item A dataset of 1,861 real-world song--MV pairs with song-side, MV-side, and teacher-inferred rationale annotations, together with a LoRA-based SFT model trained to generate global and local bridges for hierarchical MV planning.
    \item Experimental results show improvements in song-grounded visual translation, visual development, and long-range conceptual and shot-level coherence, with ablations examining the effects of bridge supervision and conditioning.
\end{itemize}

  \section{Method}
\label{sec:method}

As shown in Fig.~\ref{fig:mvweaver_pipeline}, given a song \(x\), \sysname combines a comprehensive analysis of the whole song,
hierarchical visual planning, and downstream generation. A learned
song-to-visual bridge maps song-side evidence to global and local
bridge decisions that condition the visual planning process.

\subsection{Hierarchical Music Video Agent}
\label{sec:hierarchical_agent}

\paragraph{Comprehensive song analysis}
\sysname combines song-level understanding with fine-grained lyrical,
structural, and acoustic analysis.
At the song level, Gemini 3.1 Pro summarizes the complete audio into an
eight-dimensional musical and semantic profile. At the fine-grained level,
HT-Demucs~\cite{rouard2022hybrid} isolates vocals for timestamped transcription
by FireRedASR2S~\cite{xu2026fireredasr2s}, while
SongFormer~\cite{hao2025songformer} segments musical sections. We align the
lyrics to these sections and use Librosa~\cite{mcfee2015librosa} to compute
their normalized, smoothed log-RMS energy. We denote the resulting song
representation by
\(\mathcal{S}=(s^{\mathrm{song}},\{s_i^{\mathrm{sec}}\}_{i=1}^{N})\), where
\(s^{\mathrm{song}}\) is the song-level profile and \(s_i^{\mathrm{sec}}\)
aligns the label, time range, lyrics, and energy of section \(i\).

Separately, the madmom~\cite{bock2016joint} estimates beats
and downbeats, while Librosa detects strong onsets. We merge downbeat and onset
candidates within \(0.3\) seconds into the rhythmic candidate set
\(\mathcal{C}\).
The semantic and structural analyses guide what the visual plan should express
and how it evolves, while \(\mathcal{C}\) provides temporal anchors for scene
and shot boundaries.

\paragraph{Bridge-conditioned hierarchical planning}
The Qwen3.6-27B bridge model~\cite{team2026qwen3} introduced in
Sec.~\ref{sec:bridge_learning} maps \(\mathcal{S}\) to
\(\widehat{B}=(\widehat{B}^{g},
\{\widehat{b}_j^{l}\}_{j=1}^{M})\), where \(\widehat{B}^{g}\) is the global
bridge and \(\widehat{b}_j^{l}\) is the \(j\)-th of \(M\) ordered local bridge
decisions. Here, global refers to decisions about the overall MV concept, whereas local refers to decisions about its ordered visual developments. The
planner itself is organized into MV-brief, visual-development, scene, and shot
stages. Gemini 3.1 Pro planners progressively expand these decisions as follows:
\begin{equation}
    \begin{aligned}
        z^{\mathrm{brief}} &=
        \mathcal{P}_{b}\!\left(\mathcal{S},\Pi_g(\widehat{B}^{g})\right),\\
        \left(\mathcal{R},\mathcal{Z}\right)
        &=\mathcal{P}_{d}\!\left(
        z^{\mathrm{brief}},\mathcal{S},
        \Pi_l(\{\widehat{b}_j^{l}\})\right),\\
        \mathcal{E}
        &=\mathcal{P}_{s}\!\left(\mathcal{Z},\mathcal{R},\mathcal{S}\right),\\
        \mathcal{Q}
        &=\mathcal{P}_{q}\!\left(\mathcal{E},\mathcal{R},\mathcal{S}\right).
    \end{aligned}
    \label{eq:hierarchical_planning}
\end{equation}
The MV brief planner \(\mathcal{P}_b\) turns the global bridge into an MV brief
\(z^{\mathrm{brief}}\). The visual development planner \(\mathcal{P}_d\)
combines this brief with the ordered local bridge decisions to form visual
developments \(\mathcal{Z}=\{z_j^{\mathrm{dev}}\}\) and defines reusable assets
\(\mathcal{R}\), including characters, outfits, and locations. The scene
planner \(\mathcal{P}_s\) expands each development into scenes
\(\mathcal{E}\) with asset assignments, transitions, and montage. The shot
planner \(\mathcal{P}_q\) decomposes them into renderable shots
\(\mathcal{Q}\) with timing, visual and camera instructions, and generator
routing. Scene and shot boundaries are aligned to nearby rhythmic
candidates.

\paragraph{Reference-conditioned generation}
GPT Image 2 renders the reusable assets \(\mathcal{R}\) and
reference-conditioned first frames specified by \(\mathcal{Q}\). Kling v3
animates non-singing shots, while Kling Avatar generates singing shots from
their first frames and vocal segments.

\subsection{Real-MV-Derived Song-to-Visual Bridge Dataset}
\label{sec:bridge_dataset}
Within professional MV creation, song-to-visual planning relates song evidence to visual ideas and develops these associations into concrete decisions, rather than merely illustrating the song literally.
Inspired by conceptual blending, we use an association schema to
construct a dataset of 1,861 songs paired with their real-world MVs, using
1,841 pairs for training and 20 for testing. For each pair, we derive
song-side analysis \(\mathcal{S}\) and an MV-side representation
\(\mathcal{V}\) containing its overall visual design and temporal visual
development. Gemini 3.1 Pro serves as the teacher that contrasts the two
representations and retrospectively infers the target bridge
\begin{equation}
    B^{*}=\mathcal{T}(\mathcal{S},\mathcal{V})
    =\left(B^{g*},\{b_j^{l*}\}_{j=1}^{M}\right).
    \label{eq:bridge_annotation}
\end{equation}

For each real MV, we construct video-side annotations at two levels. The global annotation describes its overall visual design, while the local annotations organize temporally related segments into visual developments.
The global bridge \(B^{g*}\)  comprises six MV design dimensions covering type, theme, summary concept, visual style, aesthetic reference, and recurring visual
motifs. Each dimension records the corresponding design decision, supporting song evidence, an association chain, and its intended effect. The local bridge represents the reference MV as a small number of ordered visual developments. Each development contains its song
evidence, visual progression, and association chain.

Each paired MV provides one concrete visual realization of its song. Beyond supervising its final visual decisions, we use a teacher model to construct five-stage association chains linking song evidence to observed MV decisions through conceptual abstraction, cross-domain mapping, and visual concept formation. This process supervision encourages the model to learn transferable
song-to-visual relations rather than reproduce isolated decisions from individual MVs. We will release the source MV links and complete audiovisual annotations.
\subsection{Bridge Learning and Agent Integration}
\label{sec:bridge_learning}

We perform parameter-efficient supervised fine-tuning (SFT) of Qwen3.6-27B
using LoRA~\cite{hu2022lora}. Given only song-side information
\(\mathcal{S}\), the model jointly predicts the two bridge levels
retrospectively inferred from the paired real MV. We use rank \(8\),
\(\alpha=16\), and train for two epochs, resulting in \(58.36\)M trainable
parameters. We optimize the standard autoregressive language-modeling loss over
the bridge targets. Joint prediction encourages the local bridge
developments to remain compatible with the global bridge concept instead of
forming independent section-wise ideas.

At inference, the trained model generates
\(\widehat{B}=f_{\mathrm{bridge}}(\mathcal{S})\). In
Eq.~(\ref{eq:hierarchical_planning}), \(\Pi_g\) and \(\Pi_l\) project the global
and local bridge decisions into \(\mathcal{P}_b\) and \(\mathcal{P}_d\),
respectively. Later stages inherit them through \(z^{\mathrm{brief}}\) and
\(\mathcal{Z}\). Song evidence and association chains serve as supervision
only and are not appended to planner prompts.

  \section{Experiments}
\label{sec:experiments}

\begin{figure*}[t]
    \centering
    \includegraphics[width=\textwidth]{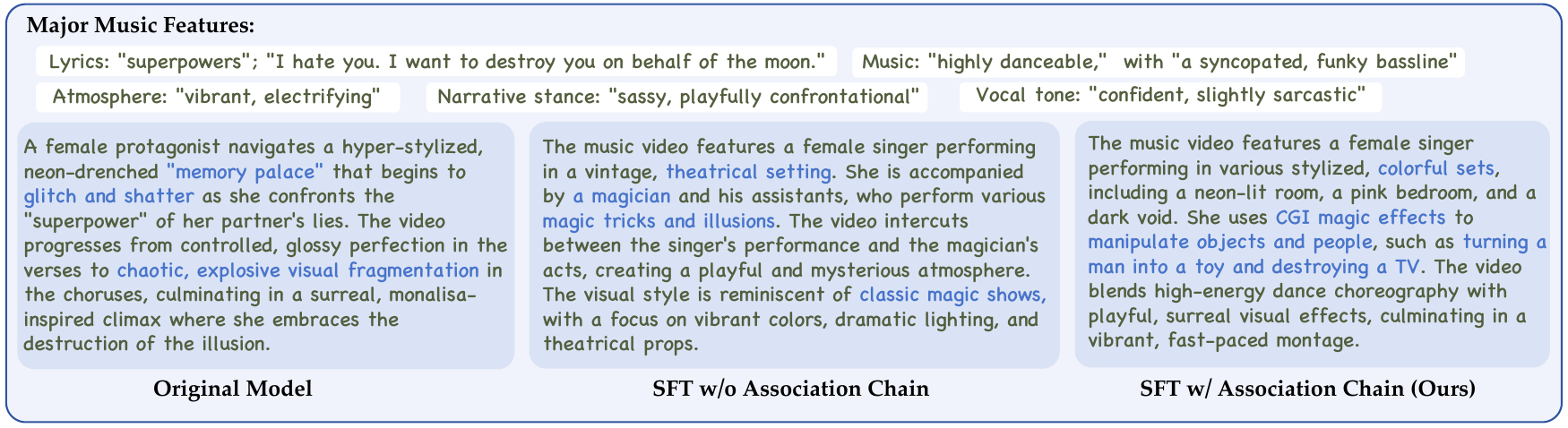}
    \caption{Qualitative comparison of global bridge outputs for G.E.M.'s
    \emph{Superpower}. All variants receive the same song-side information,
    excerpted at the top. The original model uses an abstract ``memory
    palace,'' a generic psychological metaphor that weakly reflects the song's
    playful sarcasm. SFT without the association chain follows a surface
    lexical link from ``superpower'' to magician-led stage magic, diluting the
    protagonist's confrontational agency. Full SFT instead translates these
    cues into protagonist-controlled magical actions, yielding a more
    song-specific MV premise. Due to space constraints, only the MV summary
    from each global bridge output is shown.}
    \label{fig:chain_ablation}
\end{figure*}

\paragraph{Evaluation protocol and comparisons.}
We evaluate plans for 20 test songs against the closest publicly available full-song baseline, AutoMV~\cite{tang2025automv}, and an internal \emph{w/o Bridge} ablation. For fairness, we run AutoMV with the same Gemini 3.1 Pro LLM and GPT Image 2 image generator as \sysname. The ablation retains our pipeline but removes global and
local bridge inputs. We use GPT-5.6-sol for blinded
five-point scoring.

\paragraph{Metrics.}
Following the two requirements in Sec.~\ref{sec:introduction}, we use
Interpretation Song Alignment (ISA) to assess whether a plan captures the
song's themes, emotional trajectory, and musical development without reducing
the lyrics to mechanical illustration. Grounded Directorial Effectiveness (GDE)
measures whether song-grounded ideas develop into a concrete progression with
motivated stages, a climax, and a resolution rather than unsupported
elaboration. For long-form planning, Conceptual Coherence (CC) measures whether
the full plan sustains and develops a unified visual premise, world, and
recurring motifs. Shot-to-Shot Continuity (SSC) measures whether identities,
spaces, props, actions, and relationship states remain traceable across cuts
and narrative branches, with changes motivated and prior states carried
forward or resolved. All scores range from 1 to 5. Higher is better.

\begin{table}[t]
    \caption{
    Quantitative evaluation and user preference rates. ISA and GDE assess song interpretation and grounded visual development, while CC and SSC assess long-
  form coherence. The \sysname preference rates are averaged across pairwise comparisons with all comparison methods. ``w/o Bridge'' denotes the internal ablation.
    }
    \label{tab:planning_evaluation}
    \centering
    \small
    \setlength{\tabcolsep}{2pt}
    \resizebox{\columnwidth}{!}{%
    \begin{tabular}{lccccccc}
        \toprule
        & \multicolumn{2}{c}{Song Grounding}
        & \multicolumn{2}{c}{Long-Form Coherence}
        & \multicolumn{3}{c}{User Preference (\%)} \\
        \cmidrule(lr){2-3}\cmidrule(lr){4-5}\cmidrule(lr){6-8}
        Method & ISA \(\uparrow\) & GDE \(\uparrow\)
        & CC \(\uparrow\) & SSC \(\uparrow\)
        & Interp. & Coher. & Overall \\
        \midrule
        AutoMV & 3.06 & 3.23 & 3.95 & 3.10
        & 37.65\% & 23.53\% & 30.59\% \\
        w/o Bridge & 3.82 & 3.44 & 4.25 & 3.90
        & 32.94\% & 36.47\% & 35.29\% \\
        MVWeaver & \textbf{4.08} & \textbf{3.63}
        & \textbf{4.30} & \textbf{4.10}
        & \textbf{64.71}\% &  \textbf{70.00}\%& \textbf{67.06}\% \\
        \bottomrule
    \end{tabular}
    }
\end{table}

\paragraph{Results and qualitative ablation.}
As shown in Table~\ref{tab:planning_evaluation}, \sysname consistently
outperforms AutoMV in both song-grounded translation and long-form planning
coherence. AutoMV is particularly weak on SSC because its segment-wise
generation anchors appearance but does not propagate evolving states across
shots, leading to fragmented transitions and state resets. Relative to w/o
Bridge, \sysname improves ISA and GDE, demonstrating
the benefit of the learned bridge for song grounding and visual development.
It also improves SSC, while the smaller gain in CC reflects their shared
hierarchical planner and reusable assets.
Fig.~\ref{fig:chain_ablation} illustrates that SFT alone can still rely on
surface associations, whereas association-chain supervision better preserves
song-specific stance and agency in the resulting visual premise.

\paragraph{Qualitative comparison.}
The supplementary video compares the final MVs produced by \sysname and
AutoMV. AutoMV often falls back on generic visual situations, such as a person
walking down a street, that are only weakly tied to the song. Its character
bank helps anchor appearance, but the absence of reusable location references
allows segments to drift across unrelated settings. In contrast, \sysname
develops more song-specific visual premises and combines reusable location
assets with planned montage transitions to support coherent long-form
progression.

\paragraph{User study.}
We conducted a blind A/B study on the final generated MVs. We randomly
selected ten pairs comparing \sysname with the baselines, yielding 30 questions
across three criteria. For each pair, participants selected the better MV in
song interpretation, long-range coherence, and overall effectiveness. We
collected votes from 34 participants. The preference rates are reported in the
last three columns of Table~\ref{tab:planning_evaluation}. Our method achieved
higher preference rates across all three criteria.

  \section{Conclusion}
\label{sec:conclusion}

We presented \sysname, a hierarchical agent for full-song MV generation that
combines long-form visual planning with a learned song-to-visual bridge.
We distill MV-derived song-to-visual rationales into a Qwen bridge model that
predicts global and local bridge decisions from song-side
information.
Experimental results suggest that this
supervision improves song-grounded visual development and coherence across the
generated sequence. Future work will incorporate professional editing
principles to model cross-shot camera-motion continuity more explicitly.

  \bibliographystyle{ACM-Reference-Format}
  \bibliography{MVWeaver}

  \end{document}